\documentclass[twocolumn,10pt]{article} 

\usepackage[square,numbers,sort&compress,comma]{natbib}

\usepackage{amsmath}
\usepackage{amssymb}
\usepackage{caption}
\usepackage{graphicx}
\usepackage{latexsym}
\usepackage{times}
\usepackage[pagewise]{lineno}
\usepackage{subeqnarray}
\usepackage{caption}
\usepackage{floatrow}
\usepackage{subfig}
\usepackage{float}
\usepackage{stfloats}
\usepackage{arydshln}
\usepackage{xcolor}
\usepackage{booktabs}
\makeatletter
\newcommand\rlarrows{\mathop{\operator@font \rightleftarrows}\nolimits}
\makeatother
 \def\Avec{{\mbox{\boldmath$A$}}}

 \def\Bvec{{\mbox{\boldmath$B$}}}
 
\def\Cvec{{\mbox{\boldmath$C$}}}

 \def\fvec{{\mbox{\boldmath$f$}}}

 \def\Ivec{{\mbox{\boldmath$I$}}}

 \def\pvec{{\mbox{\boldmath$p$}}}
 \def\Pvec{{\mbox{\boldmath$P$}}}
 \def\uvec{{\mbox{\boldmath$u$}}}

 \def\Rvec{{\mbox{\boldmath$R$}}}
 \def\rvec{{\mbox{\boldmath$r$}}}

 \def\xvec{{\mbox{\boldmath$x$}}}

 \def\qvec{{\mbox{\boldmath$q$}}}
 \def\0vec{{\mbox{\boldmath$0$}}}

 \def\etavec{{\mbox{\boldmath$\eta$}}}

 \def\tauvec{{\boldsymbol{\tau}}}

\topmargin - 12pt 
\renewenvironment{abstract}%
              {
               \small
               {\bfseries \abstractname}
               \par
               \vspace{10pt}
              }

\renewcommand\abstractname{Abstract}

\newcommand{\nomenclature}
              [1]
              {
               \bgroup
               \flushleft
               \small\bf
               #1
               \par
               \egroup
              }

\renewcommand{\section}
              [1]
              {
               \bgroup
               \flushleft
               \small\bf
               \refstepcounter{section}
               \arabic{section}. #1
               \par
               \egroup
              }

\renewcommand{\subsection}
              [1]
              {
               \bgroup
               \flushleft
               \small\em
               \refstepcounter{subsection}
               \arabic{section}.
               \arabic{subsection}. #1
               \par
               \egroup
              }

\renewcommand{\subsubsection}
              [1]
              {
               \bgroup
               \flushleft
               \small\em
               \refstepcounter{subsubsection}
               \arabic{section}.
               \arabic{subsection}.
               \arabic{subsubsection}. #1
               \par
               \egroup
              }

  \newcommand{\acknowledgement}
              [1]
              {
               \bgroup
               \flushleft
               \small\bf
               #1
               \par
               \egroup
              }

  \newcommand{\sectionbib}
              [1]
              {
               \bgroup
               \flushleft
               \small\bf
               #1
               \par
               \egroup
              }

\begin{document}

\title{\LARGE Coupled thermoacoustic resolvent analysis of a model two-stream coaxial combustor}

\author{{\large Lu Chen$^{a,b}$, Zheng Qiao$^{a,c}$, Yu Lv$^{a,c,*}$}\\[10pt]
        {\footnotesize \em $^a$State Key Laboratory of Nonlinear Mechanics, Institute of Mechanics, Chinese Academy of Sciences, Beijing 100080,
China}\\[-5pt]
        {\footnotesize \em $^b$Department of Mechanical Engineering, National University of Singapore, 9 Engineering Drive 1,
117575, Republic of Singapore}
        \\[-5pt]
        {\footnotesize \em $^c$School of Engineering Sciences, University of Chinese Academy of Sciences, Beijing 100049, China}\\[-5pt]
        }

\date{}


\small
\baselineskip 10pt


\twocolumn[\begin{@twocolumnfalse}
\vspace{50pt}
\maketitle
\vspace{40pt}
\rule{\textwidth}{0.5pt}
\begin{abstract} 
We derived a resolvent operator to analyze the coupled flame-acoustic effect in a model two-stream coaxial combustor. The theoretical analysis accommodates both the hydrodynamic effect of the active flame and chamber acoustic effect, as well as their coupling effect. Utilizing the coupled thermoacoustic resolvent, we are able to identify the optimal forcing of different mechanisms and their corresponding optimal responses. When this new modeling tool is applied to a two-stream coaxial model combustor, it reveals distinctly different forcing-response characteristics in the coupled system compared to those in a purely hydrodynamic linear system. Acoustic energy peaks, indicative of resonance, are observed. Additionally, it has been found that the acoustic forcing does not maintain a rank-one property. Furthermore, the modes most receptive to the flame-acoustic coupling effect can be identified using a novel sensitivity analysis approach. Interestingly, by employing the coupled thermoacoustic resolvent, we demonstrate that forcing can be designed to negate or counteract perturbations caused by the forcing of another attribute, potentially aiding in the development of effective thermoacoustic control strategies.
\end{abstract}
\vspace{10pt}
\parbox{1.0\textwidth}{\footnotesize {\em Keywords:} thermoacoustic; resolvent; optimal forcing; optimal response}
\rule{\textwidth}{0.5pt}
\vspace{180pt}

*Corresponding author\\
Email: lvyu@imech.ac.cn (Y. Lv)

\end{@twocolumnfalse}] 

\clearpage

\twocolumn[\begin{@twocolumnfalse}

\centerline{\bf Information for Colloquium Chairs and Cochairs, Editors, and Reviewers}

\vspace{20pt}


\vspace{20pt}

{\bf 1) Novelty and Significance Statement}
\vspace{10pt}


\vspace{10pt}

This study proposes a coupled thermoacoustic (CTA) resolvent analysis framework, aiming to gain new insights into the forcing-response relations of closed-loop thermoacoustic systems. Diverging from prior research, this approach \textit{shifts the focus to the combustor rather than the flame}. It allows for the exploration of various forcing mechanisms and their corresponding responses. Intriguingly, the CTA resolvent can be utilized to design input forcings that cancel or counteract the responses led by an existing forcing, This novel theoretical approach has the potential as an efficient tool for thermoacoustic control.

\vspace{20pt} 

{\bf 2) Author Contributions}
\vspace{10pt}


\begin{itemize}

  \item{LC: designed research; performed research; analyzed data; wrote the paper.}

  \item{ZQ: designed research; performed research; analyzed data; wrote the paper.}

  \item{YL: designed research; performed research; analyzed data; wrote the paper.}

\end{itemize}

\vspace{10pt}

{\bf 3) Authors' Preference and Justification for Mode of Presentation at the Symposium}
\vspace{10pt} 


\vspace{10pt}

The authors prefer {\bf OPP} presentation at the Symposium, for the following reasons:

\begin{itemize}

  \item{This paper proposes a novel thermoacoustic analysis method that is different from other related works.}

  \item{This work improves current tools' predictive capabilities for analyzing thermoacoustic instability.}

  \item{The presentation is focusing on the prediction results, incorporating only the essential background information.}
  
\end{itemize}

\end{@twocolumnfalse}] 


\clearpage


\section{Introduction} \addvspace{10pt}

Classical thermoacoustic studies on propulsion systems often focus on the stability analysis to identify the unstable eigenmodes. It has been recognized that the thermoacoustic system also exhibits the transient-growth or pseudo-resonance due to the non-normality nature of the linear system. The related short-term behaviors may equally cause detrimental consequences to combustion systems. To effectively analyze such short-term behaviors, the analysis based on the resolvent operator has gained increasing attention. The resolvent analysis directly reveals the amplification property of a non-normal linear system to external perturbations. In the fields of thermoacoustics, the resolvent analysis was used to construct the relation between velocity perturbation (input) and heat-release fluctuation (output) for a laminar M-shaped premixed flame~\cite{blanchard2015response} and for a laminar premixed slot flame~\cite{wang2022linear}. More recently, the resolvent operator was leveraged to study the effects of swirl on the flame frequency response~\cite{skene2019adjoint}.  The efforts of exploiting the capability of the resolvent operator were also extended to the studies of turbulent flames. The flame responses to external forcings were investigated in terms of both passive~\cite{casel2022resolvent} and active~\cite{kaiser2023modelling} jet flames. It has also been taken to predict the flow response in turbulent swirling flames to acoustic perturbations~\cite{kaiser2019prediction}. 

In the previous resolvent analyses, the focus is often placed on the hydrodynamics in the flame region at the low Mach number limit, with the goal of finding the forcing under which heat-release rate fluctuations are maximized\cite{sayadi2021frequency,wang2022linear,kaiser2023modelling}. However, as such the flame is considered as a decoupled open-loop system while the chamber effect that delineates acoustic reflection and resonance is ignored. Although this technique could be useful for the construction of flame transfer functions (FTF), it does not provide a complete picture that describes the coupling between flame and acoustics~\cite{ihme2017combustion}. More importantly, the optimal forcing structure identified in those resolvent analyses does not necessarily play the same role as that in the coupled flame-acoustic system. These research issues inspire us to derive a coupled resolvent analysis framework which can be used to construct the forcing-response relations of thermoacoustic system. Toward characterizing the thermoacoustic modality in the coupled systems, the studies~\cite{magri2014global, orchini2016linear} invoked the compact flame assumption and deemed acoustic velocity at the reference point as a uniform perturbation to the flame. A multi-scale expansion approach was also proposed to model the force between the hydrodynamic and the acoustic field~\cite{magri2017multiple}. Some of these coupling strategies will be leveraged in our theoretical framework.

The objective of this study is to construct a coupled thermoacoustic (CTA) resolvent analysis framework, with the aim of gaining new insights into the forcing-response relations of closed-loop thermoacoustic systems. The key difference from the previous studies is that \textit{the focus is on the combustor instead of the flame}. The forcing mechanism of different attributes can be considered and the corresponding responses can be identified. Intriguingly, by exploiting the (CTA) resolvent, we are able to design input forcings that cancel or counteract the responses led by an existing forcing. The potential of the new theoretical method as a tool for effective thermoacoustic control will be demonstrated.

The rest of the paper is structured as follows. In section 2, the governing equations and framework of coupled thermoacoustic resolvent analysis are derived. In section 3, the configuration of the model combustor and computation setup are described. Some Numerical techniques are introduced in Section 4. The results of the coupled thermoacoustic resolvent analysis are presented and discussed in section 5, as well as the comparison with baseline results. Finally, in section 6, the main conclusions are summarised.

\begin{figure}[htbp!]
\centering
 \includegraphics[width=\textwidth]{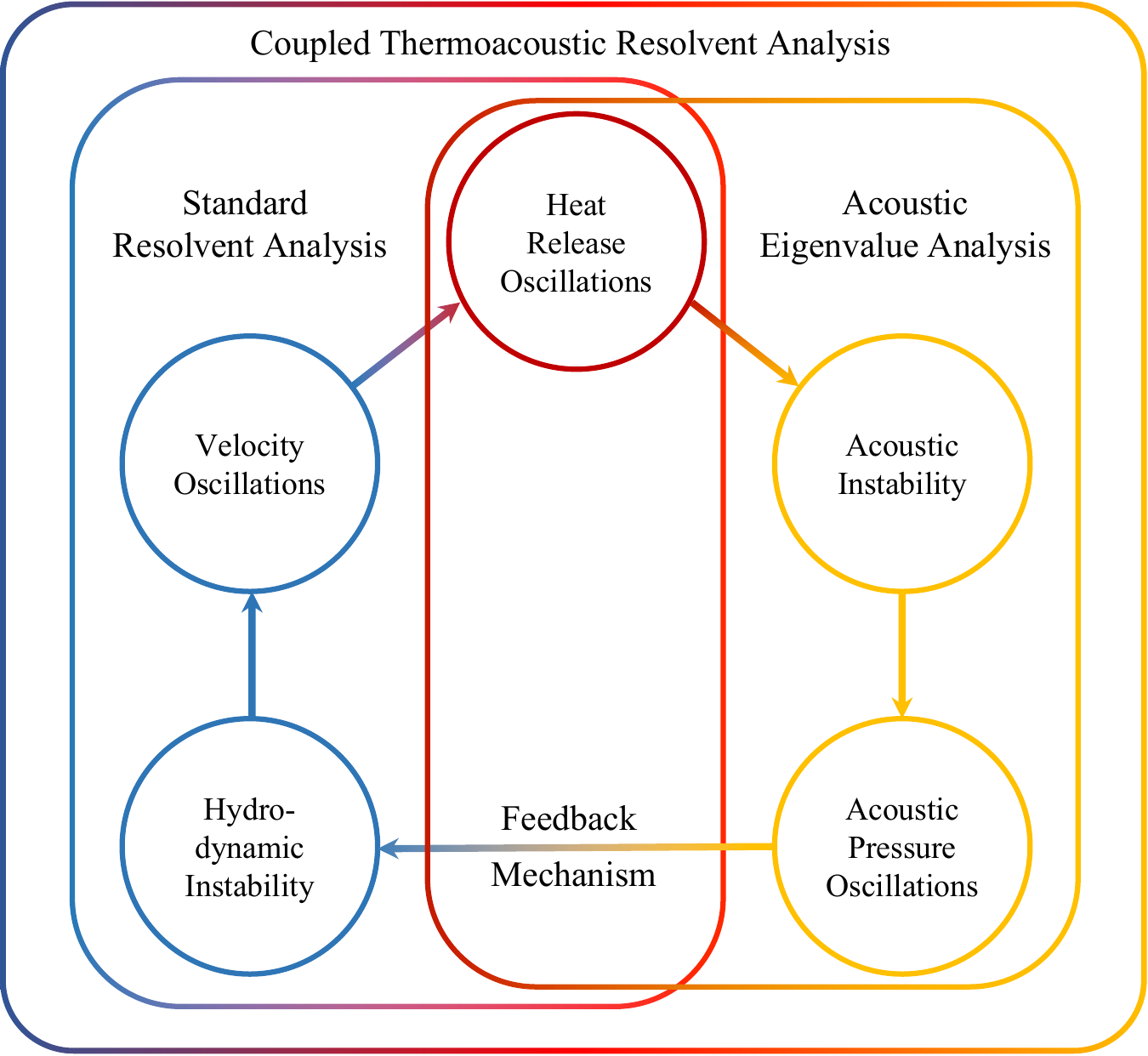}
\caption{Illustration of coupled thermoacoustic resolvent analysis.}
\label{workflow_fig}
\end{figure}
\vspace{-10pt}

\section{Mathematical formulations} \addvspace{10pt}
The low-Mach approximation is employed to simplify the representation of combustion physics since this condition is commonly fulfilled for gas-turbine combustors. The hydrodynamic effects are preserved in the flame model and examined from the linearized reactive flow equations. The acoustic propagation and reflection in the whole domain are described using the Helmholtz equation. A non-compact source term is added to the Helmholtz equation to model the induced heat release response of the flame. The acoustics are coupled to the flame/flow momentum equation by matching the velocity disturbance at a given location. A similar coupling strategy was employed previously in the studies~\cite{magri2014global, orchini2016linear}, where however the hydrodynamic modality was not explicitly addressed.

\subsection{Governing equations of flame} \addvspace{10pt}
The combustion is governed by the reactive flow equations. Here we consider a Burke-Schumann flame case by assuming an infinitely fast chemistry model. The governing equations are written as: 
\begin{subequations}
\label{govern_equ} 
\begin{align}
&\partial_t \rho +\nabla \cdot\left(\rho \vec{\uvec}\right)  = 0\;,\\
&\partial_t \left( \rho \vec{\uvec} \right) +\nabla \cdot\left(\rho \vec{\uvec} \vec{\uvec}^T\right)  = -\nabla p + \nabla  \cdot \tauvec\;, \\
&\partial_t \left(\rho Z\right) +\nabla \cdot\left(\rho \vec{\uvec} Z\right)  =  \nabla  \cdot \left(\rho D \nabla Z\right)  \;, \\
&\tauvec = \mu \left(\left(\nabla \vec{\uvec}\right)+ \left(\nabla \vec{\uvec}\right)^{T}\right) - \frac{2}{3} \mu \left(\nabla \cdot \vec{\uvec}\right)\vec{\vec{\Ivec}}\;, 
\end{align}
\end{subequations}
in which $\rho$ is the density, $\vec{\uvec}$ is the velocity, $\tauvec$ is the viscous stress tensor, $p$ is the hydrodynamic pressure  and $Z$ denotes the mixture fraction. Given the fast chemistry assumption, the species concentration $Y$, temperature $T$, and transport properties (including mass diffusivity $D$ and dynamic viscosity $\mu$) are functions of $Z$ and can be pre-tabulated with respect to $Z$.

\subsection{\label{model_flame}Linear model of combustion}
To examine the effects of convective perturbation on the flame, we therefore treat the flame as an amplifier that takes certain external forcing as input and gives back heat-release disturbance as the output. The input-output relations are constructed by considering the resolvent of the linearized reactive flow equations. The combustion linear system is derived from Eq.~(\ref{govern_equ}) and can be cast to the form:
\begin{equation}
\label{eqRstand}
\Bvec_{h}^0\frac{\partial{\qvec}_{h}^{\prime}}{\partial t} = \Avec_{h}^0{\qvec}_{h}^{\prime}+  {\fvec}_{h}^{\prime}
\end{equation}
where $\Avec_{h}^0$ and $\Bvec_{h}^0$ are matrices evaluated around the steady-state flame solution, $\qvec^0$. The $\qvec^\prime_h$ denotes a set of independent variables of the linearized system, for instance, $\qvec^\prime_h = (\vec{\uvec}^\prime, p^\prime, Z^\prime)_h$, which is solved in the flame region; and herein the forcing ${\fvec}_{h}^{\prime}$ denotes certain hydrodynamic or compositional perturbations. Once the modal transformation $({\qvec}^\prime,~{\fvec}^\prime) = (\widehat{\qvec},~\widehat{\fvec})\exp(i\omega t)$ is introduced, the linear system is transformed to
\begin{equation}
\label{acoustic_flame_model}
\widehat{{{\qvec}}}_h = \Rvec_h \widehat{{\fvec}}_{h},~  \Rvec_h(\omega)=(i\omega \Bvec_{h}^0 - \Avec_{h}^0)^{-1}, 
\end{equation}
where $\Rvec_h$ is the resolvent operator providing an input-output mapping of the linear system. Through the singular value decomposition of $\Rvec_h$, we are able to seek the optimal forcing with which the energy gain of $||\widehat{\qvec}_h||^2 / ||\widehat{\fvec}_h||^2$ is maximized. Note that for combustion problem one often concerns the energy norm of heat release fluctuation, $||\widehat{Q}||^2$, in which $\widehat{Q}$ can be expressed as 
\begin{equation}
 \begin{aligned}
\widehat{Q} & =  \rho^0 c_p^0 (i\omega \widehat{T}_h  +  \vec{\uvec}^0 \cdot \nabla \widehat{T}_h   +   \widehat{\vec{\uvec}}_h \cdot \nabla T^0  ) \\
 & +  \left( \widehat{\rho}_h c_p^0 + \rho^0 \widehat{c_p}_h \right) \vec{\uvec}^0 \cdot \nabla T^0\;,   
  \end{aligned}
\end{equation}
where $c_p$ is the heat capacity at constant pressure. For Burke-Schumann flames we may write $\widehat{Q}= \Cvec_h^0(\omega) \widehat{\qvec}_h$ since $\widehat{T}$, $\widehat{\rho}$ and $\widehat{c_p}$ are all functions of $\widehat{Z}$. Note that in this study, $\widehat{Q}$ is defined as a local quantity with spatial dependency, not an integral quantity as employed for compact flame analyses~\cite{illingworth2013finding, magri2014global}. Since the resolvent operator directly relates the heat-release perturbation with the input velocity disturbance, hence it has been used to effectively estimate FTFs for thermoacoustic analysis~\cite{blanchard2015response, wang2022linear}. 

\begin{table}
\begin{tabular}{lll}
  \toprule
   &  $||\widehat{\fvec}||^2$ & $||\widehat{\qvec}||^2$   \\
  \midrule
  $\widehat{\fvec}_{h,u_x}$    & $\int_{\Omega_{h}} |\widehat{\fvec}_{h,u_x}|^{2} rdrdx$  & $\int_{\Omega_{h}} |\widehat{Q}|^{2} rdrdx$ \\
 $\widehat{\fvec}_{h,Z}$  & $\int_{\Omega_{h}} |\widehat{\fvec}_{h,Z}|^{2} rdrdx$   & $\int_{\Omega_{h}} |\widehat{Q}|^{2} rdrdx$ \\
  \bottomrule
\end{tabular}
\caption{Energy norm to $\Rvec_h$}
\end{table}
\begin{table}
\begin{tabular}{lll}
  \toprule
   &  $||\widehat{\fvec}||^2$ & $||\widehat{\qvec}||^2$   \\
  \midrule
  $\widehat{\fvec}_{h,u_x}$    & $\int_{\Omega_{h}} |\widehat{\fvec}_{h,u_x}|^{2} rdrdx$  & $\int_{\Omega} \frac{{\hat{p}_{a}}^{2}}{\gamma p_{0}} rdrdx$ \\
 $\widehat{\fvec}_{h,Z}$  & $\int_{\Omega_{h}} |\widehat{\fvec}_{h,Z}|^{2} rdrdx$   & $\int_{\Omega} \frac{{\hat{p}_{a}}^{2}}{\gamma p_{0}} rdrdx$ \\
 $\widehat{\fvec}_{a}$  & $\int_{\Omega} |\widehat{\fvec}_{a}|^{2} rdrdx$   & $\int_{\Omega} \frac{{\hat{p}_{a}}^{2}}{\gamma p_{0}} rdrdx$ \\ 
  \bottomrule
\end{tabular}
\caption{Energy norm to $\Rvec_{cta}$}
\end{table}

\subsection{Acoustic model} \addvspace{10pt}
The acoustics in the combustor are described using the Helmholtz equation with a source term that represents the heat release of the flame. This model preserves the dominant source of sound generation from the combustion region and is sensible for the low-Mach regime. It reads:  
\begin{equation}
\label{acoustic_model}
\frac{1}{\gamma_0 p_{0}} \frac{\partial^{2} p_a^{\prime}}{\partial t^{2}} - \nabla \cdot(\frac{1}{\rho_{0}}\nabla p^{\prime}_a) = \frac{\gamma_0 -1}{\gamma_0 p_{0}}\frac{\partial{Q^{\prime}}}{\partial t}  \;,
\end{equation}
where $p_{0}$ and $\rho_{0}$ are the pressure and density of the steady-state solution and $\gamma_0$ is the adiabatic index. We introduce the solution variable $\widehat{\qvec}_a = (\widehat{\pvec}_a, \widehat{\rvec}_a)$ with $\widehat{\rvec}_a = i \omega \widehat{\pvec}_a$ and perform  the modal transformation as in Sec.~\ref{model_flame}. Equation~(\ref{acoustic_model}) can be recast to the linear form
\begin{equation}
\label{acoustic_linear_model}
 \Avec_a^0  \widehat{\qvec}_a = i\omega \Bvec_a^0 \widehat{\qvec}_a - i \omega \Cvec_a^0 \widehat{Q} -   \widehat{\fvec}_a  \;,
\end{equation}
with the matrices $\Avec_a^0$, $\Bvec_a^0$ and $\Cvec_a^0$ to be evaluated from the steady-state flame solution. The additional acoustic forcing $\widehat{\fvec}_a$ is introduced to the right-hand side of Eq.~(\ref{acoustic_linear_model}) to examine the transient growth of the system. In a typical engine configuration, the forcings of such kind may originate from turbomachinery parts or arise from structural vibration. Note that the acoustic velocity perturbation $\vec{\uvec}_a^\prime$, although not directly solved, can be derived from the pressure field. 

\subsection{Coupled thermoacoustic resolvent analysis} \addvspace{10pt}
From the physical view of compressible flow, the flame and acoustic effects are always coupled in actual combustors, which potentially indicates thermoacoustic instability. To investigate the coupling effect, we follow the commonly used approach~\cite{silva2013combining, salas2013physical, motheau2014mixed}. Specifically, the acoustic velocity perturbation at a reference location is selected as the input to interrogate the flame model. The reference location is typically close to and upstream of the flame. This treatment may be translated as a mathematical expression: 
\begin{equation}
\label{eq_match}
\vec{\uvec}_h^\prime(\xvec) = \vec{\uvec}_a^\prime(\xvec),~~ \text{for}~~\xvec \in \Gamma_{\text{ref}} \;,
\end{equation}
Here $\Gamma_{\text{ref}}$ denotes the location where this matching condition is enforced. For notational convenience, we may augment Eq.~(\ref{eq_match}) in terms of solution variables:
\begin{equation}
\label{eq_match_aug}
\Pvec_h(\xvec - \xvec_{\text{ref}} ) \widehat{\qvec}_h = \Pvec_a (\xvec - \xvec_{\text{ref}} ) \widehat{\qvec}_a,\;
\end{equation}
where $\Pvec$ plays a similar role as a delta function. Combining Eqs.~(\ref{acoustic_flame_model}), (\ref{acoustic_linear_model}) and (\ref{eq_match_aug}), we obtain the following coupled model relation: 
\begin{equation}
\label{eqn:reso_cta}
\hspace{-2mm}
\resizebox{0.42\textwidth}{!}{$
\begin{bmatrix}
 \widehat{\qvec}_h  \\  \widehat{\qvec}_a \\ \etavec
\end{bmatrix} \hspace{-1mm}= \hspace{-1mm}  \underbrace{
 \begin{bmatrix} i \omega\Bvec_h^0 - \Avec_{h}^0  &\hspace{-2mm} \mathbf{0} & \Pvec_h^{T} \\  i\omega \Cvec_{a}^0\Cvec_h^0 & \hspace{-2mm} i \omega \Bvec_{a}^0  - \Avec_{a}^0    &   \mathbf{0}   \\
 \Pvec_h   &   -\Pvec_a   & \mathbf{0} 
\end{bmatrix}^{-1} }_{\Rvec_{cta}}  \hspace{-1mm}
\begin{bmatrix}
\widehat{\fvec}_{h} \\ \widehat{\fvec}_a   \\  \mathbf{0} 
\end{bmatrix}$} ,  \hspace{-1mm}
\end{equation}
where $\etavec$ is simply an auxiliary variable. $\Rvec_{cta}$ denotes the {$\mathit{c}$}oupled {$\mathit{t}$}hermo{$\mathit{a}$}coustic resolvent analysis operator. For thermoacoustic problem, the output response of interest is the energy of acoustic pressure, defined as
\begin{equation}
E_{a}=\frac{1}{2}\int_{\Omega}\frac{{\hat{p}_{a}}^{2}}{\gamma p_{0}} rdrdx 
\end{equation}
of which the integration is performed for the whole domain of the combustor, $\Omega$. The optimal forcing that leads to the maximum of $E_a$ is directly relevant to thermoacoustic prediction and control. Two types of input forcing, which correspond to distinct physical mechanisms, are preserved to examine the non-normal behavior of this two-way coupled system.

\section{Combustor configuration and computational setup} \addvspace{10pt}

\begin{figure}[htpb!]
\centering
\includegraphics[width=\textwidth]{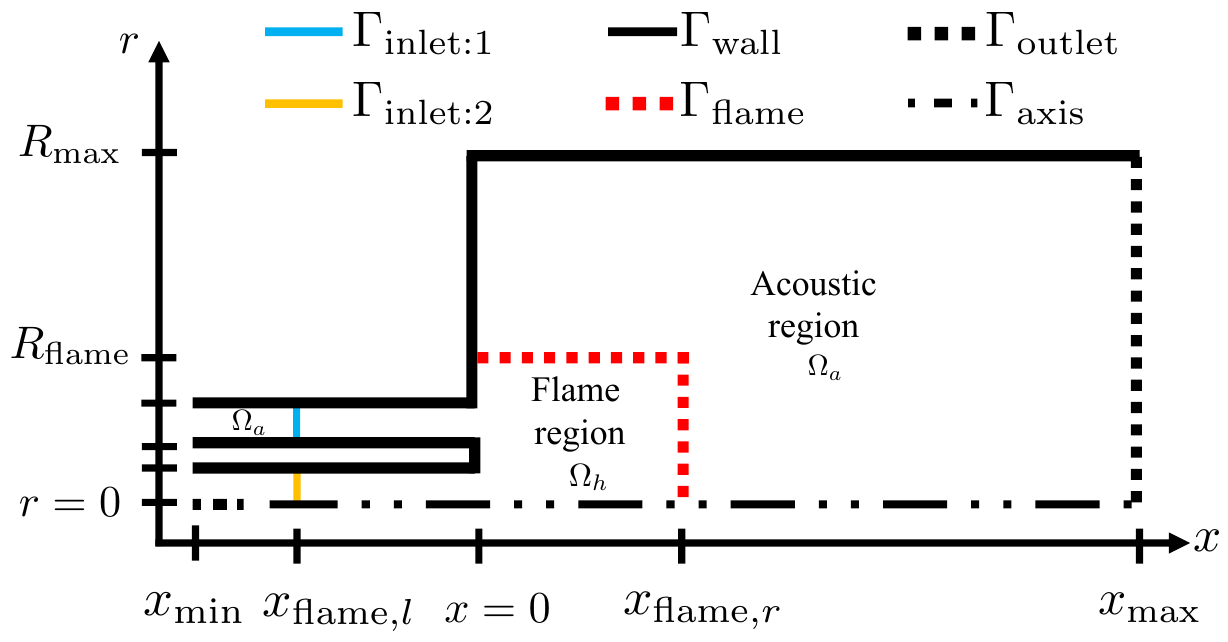}
\caption{Illustration of the combustor configuration and the computational domain.}
\label{Config_fig}
\end{figure}

Figure~\ref{Config_fig} shows the configuration of the axisymmetric model combustor considered in this study. The combustor facilitates a two-stream coaxial diffusion flame. The inner and outer nozzles issue the fuel and oxidizer, respectively. The inner nozzle diameter is $d$ while the width of the outer one is $0.4d$. A small gap of $0.1d$ separates these two nozzles. The combustor domain extends to $x \in [-50d,~240d]$ along the axial direction while keeping a height of $20d$. As discussed in the previous section, the flame linear analysis is performed on a subdomain, $\Omega_h$, where combustion occurs. $\Omega_h$ is encompassed by the boundaries (interfaces) $\Gamma_{\text{inlet}}$ and $\Gamma_{\text{flame}}$. $\Gamma_{\text{inlet}}$  is located at $x_{\text{flame,} l} = -5d$, while $\Gamma_{\text{flame}}$ is defined with $x_{\text{flame,} r} = 160d$ and $R_{\text{flame}}= 8d$. The dimension of $\Omega_h$ is purposely chosen so that the quantitative results of liner analysis are adequately captured within $\Omega_h$. 

The nonlinear simulation of reactive flow is conducted on the whole domain to generate the baseflow solution, $\qvec^0$. In this study, we are specifically interested in the oxy-rich carbon-free combustion technology. Therefore, the fuel and oxidizer are selected to hydrogen and oxygen, respectively, and issued through the nozzles with a fixed velocity ratio of $U_1 / U_2 = 0.2$. The ambient temperature $T_1$ and pressure are $300$ K and $1$ atm. Under this operating condition, the Reynolds numbers corresponding to the fuel and oxidizer streams are $200$ and $625$, respectively. The characteristic Mach number is about $0.03$.

Special attention is paid to the boundary conditions for resolvent analysis. The settings are similar to those in a recent study~\cite{douglas2023flash}. On $\Gamma_{\text{axis}}$ ($m=0$): Neumann condition for $u_{x}^\prime$, $p^\prime$ and $Z^\prime$ while Dirichlet condition for $u_{r}^\prime$ and $u_{\theta}^\prime$. On $\Gamma_{\text{outlet}}$ or $\Gamma_{\text{flame}}$: stress-free condition for $\vec{\uvec}^\prime$ while Neunmann condition for the other variables. On $\Gamma_{\text{wall}}$: no-slip conditions for velocity and Neunmann condition for the other variables. On $\Gamma_{\text{inlet}}$, the $\vec{\uvec}^\prime$ is zeroed out for standard resolvent while the matching condition Eq.~\ref{eq_match_aug} is enforced for coupled resolvent analysis. At the combustor inlet, $\qvec^\prime$ are zeroed out.

\section{Numerical techniques} \addvspace{10pt}
The governing equations for both baseflow calculation and resolvent analysis are discretized with a finite element method in the open-source software--FreeFem++~\cite{hecht2012new}. The P2 elements are used for the velocity components, and the P1 elements are used for the pressure and mixture fraction. The mesh for the whole domain comprises about 160,000 triangles, resulting in 1.6 million degrees of freedom for Eq.~(\ref{eqn:reso_cta}). 

The baseflow is obtained by solving a steady-state problem with the Newton iteration method. The resolvent analysis is conducted as a singular value problem to optimize the energy of the response field with the choice of quadratic norms:
\begin{equation}
M_{q,cta}^{1/2} \Rvec_{cta}L_{cta}M_{f,cta}^{-1/2}=X_{cta}\Sigma_{cta}Y_{cta}^{*}
\end{equation}
Here $M_{f,cta}$ and $M_{q,cta}$ are the weighting matrices to measure the energy of forcing and response mode. $L_{cta}$ has only $0$ and $1$ as elements to restrict the dimension of the linear system. The optimal and sub-optimal modes are contained in the matrices $X_{cta}$ and $Y_{cta}$, which are ordered by their corresponding energy gains $\sigma_{cta, i}^{2}$ in $\Sigma_{cta}$. This singular value decomposition is recast as an eigenvalue problem and solved by the implicitly restarted Arnoldi method. The matrix inversions in the numerical procedure are carried out with MUMPS and SuperLU provided by PETSc library~\cite{balay1998petsc}. 

\section{Results and Discussion} \addvspace{10pt}
\subsection{Base flow}
\begin{figure}[htpb!]
\centering
\includegraphics[width=\textwidth]{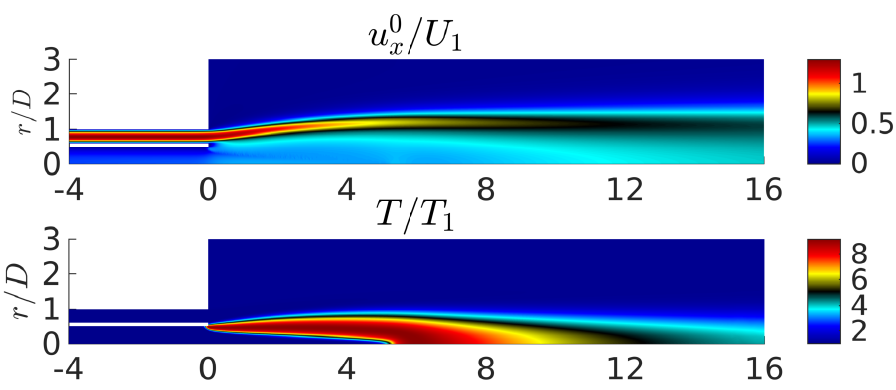}
\caption{Steady base flow for the axial velocity (top) and temperature (bottom).}
\label{baseflow}
\end{figure}

\addvspace{10pt}
The steady axisymmetric base flow is shown in Fig.~\ref{baseflow} for the axial velocity on the top and temperature on the bottom.

\vspace{-10pt}
\subsection{Baseline acoustic modes}
\begin{figure}[htpb!]
\centering
\includegraphics[width=\textwidth]{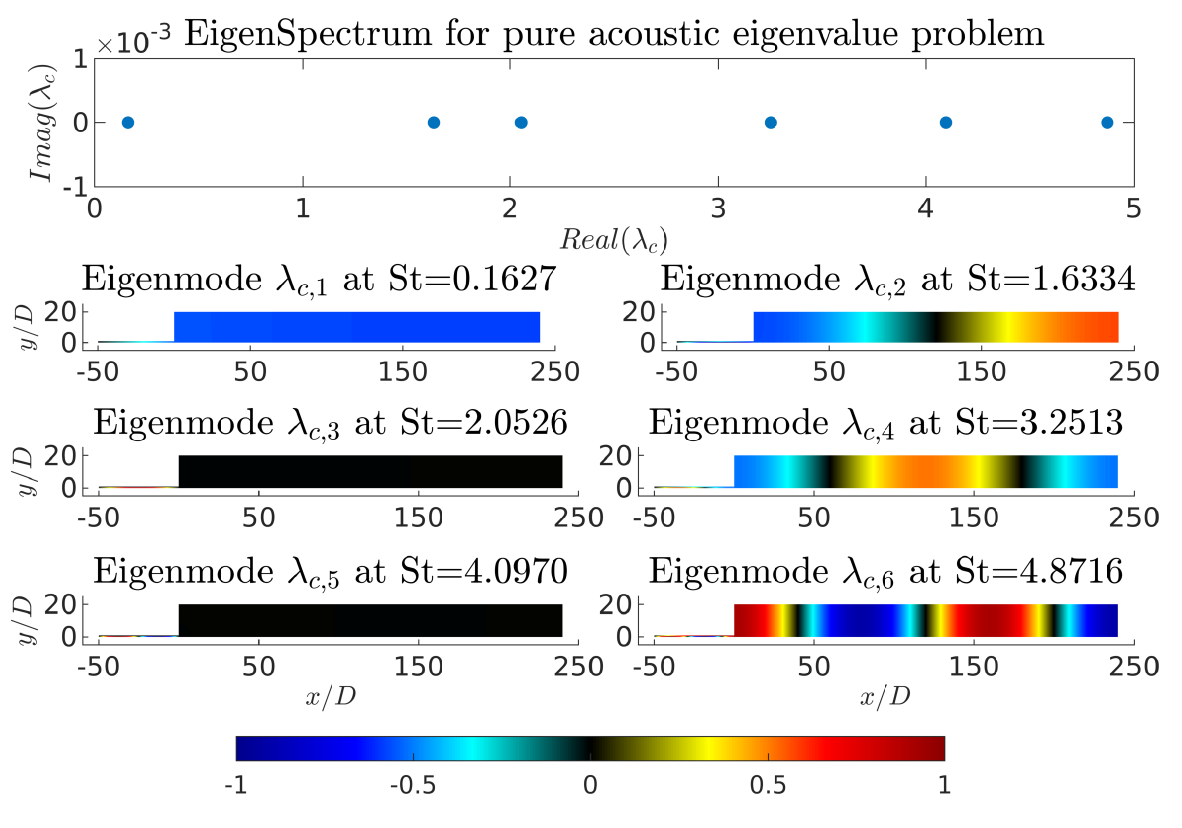}
\caption{Acoustic eigenspectrum and eigenmodes.}
\label{pureAC}
\end{figure}

\addvspace{10pt}
The acoustic eigenvalue problem with passive flame is conducted first as the baseline results for the coupled thermoacoustic resolvent analysis. The first six acoustic eigenmodes are shown in Fig.~\ref{pureAC}. Here, the acoustic eigenfrequencies $\lambda_{c}$ are the natural frequencies of the model combustor, and the $St=\omega d/U_{1}$ is the Strouhal number. The eigenmodes $3$ and $5$ are primarily concentrated in the inner nozzle, while the remaining eigenmodes are distributed throughout the combustor and may strongly interact with the flame.

\begin{figure*}[b!]
\centering
\includegraphics[width=\textwidth]{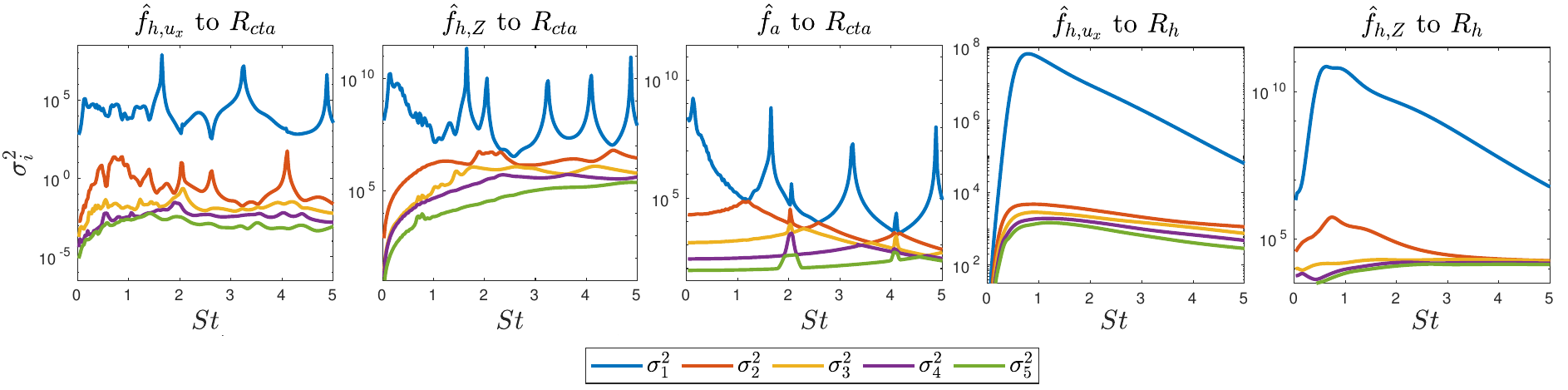}
\caption{Resolvent gains of five leading resolvent modes for $\Rvec_{cta}$ and $\Rvec_{h}$.}
\label{norm}
\end{figure*}
\vspace{-10pt}

\subsection{Resolvent analysis}\addvspace{10pt}
The leading five resolvent gains versus different Strouhal numbers are shown in Fig.~\ref{norm} for both $\Rvec_h$ and $\Rvec_{cta}$ to different components in $\widehat{{\fvec}}_{h}$ and $\widehat{{\fvec}}_{a}$. For the $\Rvec_{cta}$, several peak values exist in the optimal gain curves while only one peak value for the $\Rvec_h$.

\begin{figure}
{\begin{subfloatrow}[1]
\sidesubfloat[]{\includegraphics[width=0.95\textwidth]{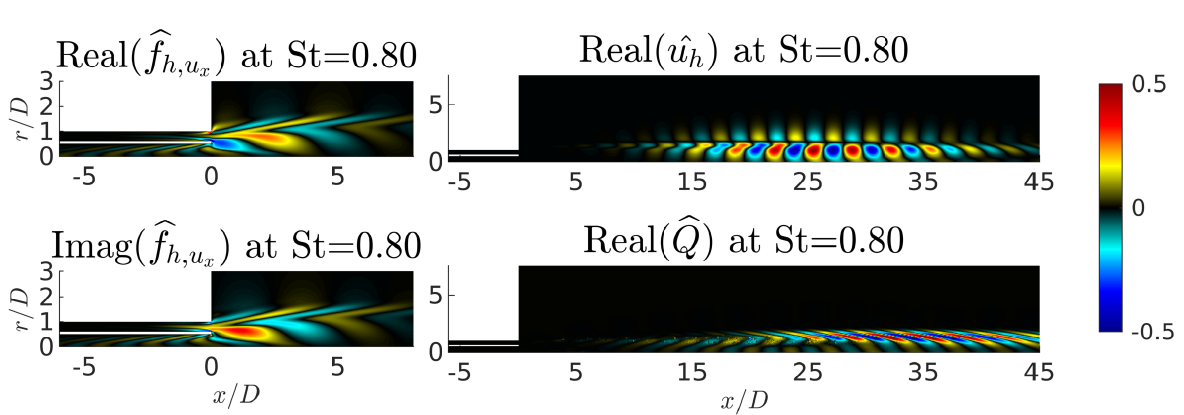}\label{fu_f_res}}%
\end{subfloatrow}}
{\begin{subfloatrow}[1]
\sidesubfloat[]{\includegraphics[width=0.95\textwidth]{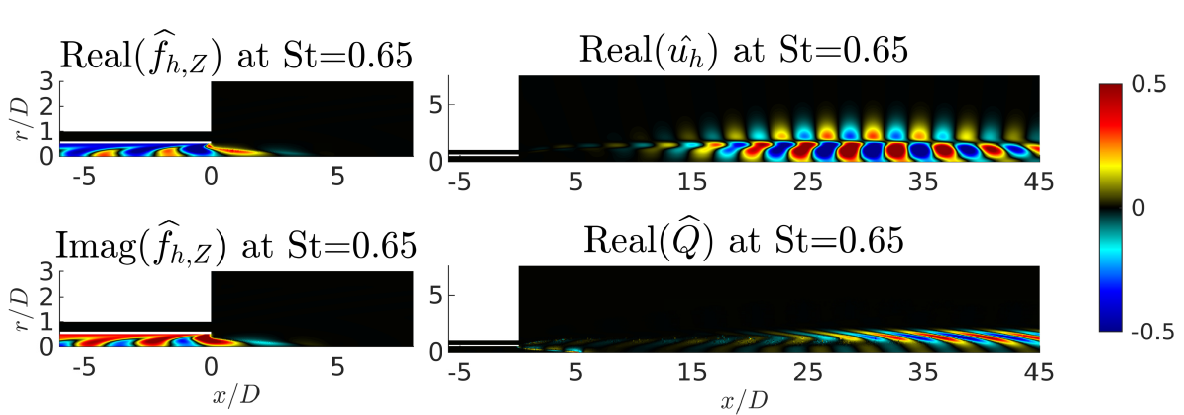}\label{fz_f_res}}%
\end{subfloatrow}}
\caption{Optimal modes of $\Rvec_{h}$. (a) Optimal forcing (left) and response (right) mode of  $\Rvec_{h}$ with respect to $\widehat{\fvec}_{h,u_{x}}$ at $St=0.80$. (b)  Optimal forcing (left) and response (right) mode of  $\Rvec_{h}$ with respect to $\widehat{\fvec}_{h,Z}$ at $St=0.65$. All fields are normalized by their maximum value.}
\end{figure}

\begin{figure}[htpb!]
\centering
\includegraphics[width=\textwidth]{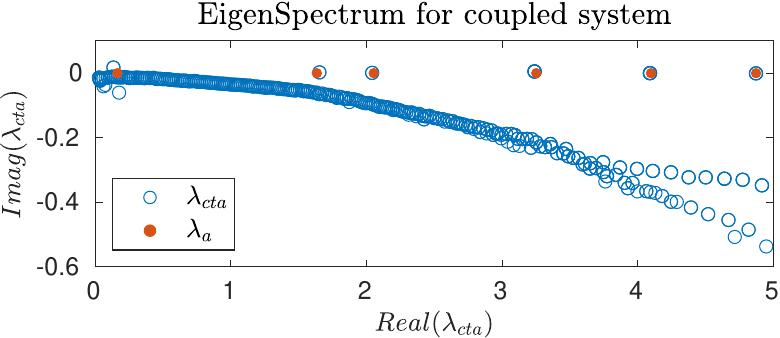}
\caption{Coupled thermoacoustic eigenspectrum.}
\label{eigenTA}
\end{figure}

No acoustic feedback is introduced in the standard resolvent analysis for $\Rvec_h$, and the flame is regarded as an energy amplifier for solely hydrodynamic perturbations. It exhibits rank-$1$ property with optimal gain $\sigma_{1}^{2}$ greatly larger than other suboptimal gains. There is only one peak value in the optimal gain curves located at $St=0.80$ for $\widehat{{\fvec}}_{h,u_{x}}$ in the streamwise momentum equation, and at $St=0.65$ for $\widehat{{\fvec}}_{h,Z}$ in the mixture fraction equation. The optimal forcing and response mode for $\widehat{{\fvec}}_{h,u_{x}}$ are shown in the Fig.~\ref{fu_f_res}. The spatially developing structure with its maximum amplitude located in the mixing layer is tilted by the mean shear, and this can be attributed to the Kelvin-Helmholtz and Orr mechanisms in the previous research \cite{casel2022resolvent}. For the $\widehat{{\fvec}}_{h,Z}$ shown in the Fig.~\ref{fz_f_res}, the optimal forcing structure representing component perturbations is mainly confined in the inner nozzle upstream of the flame and is convected by the mean flow to the flame front. Compared to the responding field of momentum perturbations shown in Fig.~\ref{fu_f_res}, here the diffusion flame is directly disturbed by the component perturbations, and the $\widehat{Q}$ is strong both near the flame front due to the chemical reactions and downstream of the flame due to the mean flow convection.
In the coupled resolvent analysis for $\Rvec_{cta}$, several peak values in the gain curves are close to the acoustic eigenfrequencies $\lambda_{c}$. Here, the eigenspectrum $\lambda_{cta}$ for the coupled system is evaluated and shown in Fig.~\ref{eigenTA}. The flame-acoustic coupling introduced in $\Rvec_{cta}$ not only combines the acoustic-related and hydrodynamic-related eigenmodes, but also leads to the shift of $\lambda_{c}$. The resolvent norm could be expanded with its eigenvalues $\lambda_{cta}$ and eigenvectors $V_{cta}$ as:
\begin{equation}
\Vert R_{cta}(\omega) \Vert \leq \underbrace{\Vert V_{cta}^{-1}\Vert \Vert V_{cta}\Vert}_{\text{pseudo-resonance}}  
\underbrace{\frac{1}{dist(i\omega,\lambda_{cta})}}_{\text{resonance}}
\end{equation}
Here the resonance term is related to the strong peaks in the gain curves of $\Rvec_{cta}$, especially in the case of $\widehat{\fvec}_a$. The pseudo-resonance term originates from the non-normal nature of both the thermoacoustic system and its hydrodynamic part. The amplification induced by pseudo-resonance can be observed from the small peaks in the gain curves of $\widehat{\fvec}_{h,u_{x}}$ around $St$ between $0.2$ and $1.5$. 

\begin{figure}
{\begin{subfloatrow}[1]
\sidesubfloat[]{\includegraphics[width=0.95\textwidth]{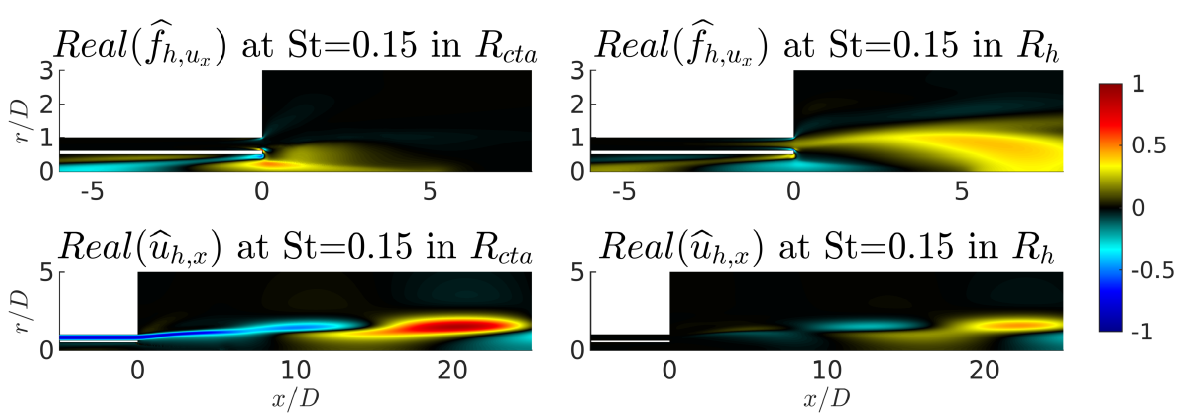}\label{compare_fu_p15}}%
\end{subfloatrow}}
{\begin{subfloatrow}[1]
\sidesubfloat[]{\includegraphics[width=0.95\textwidth]{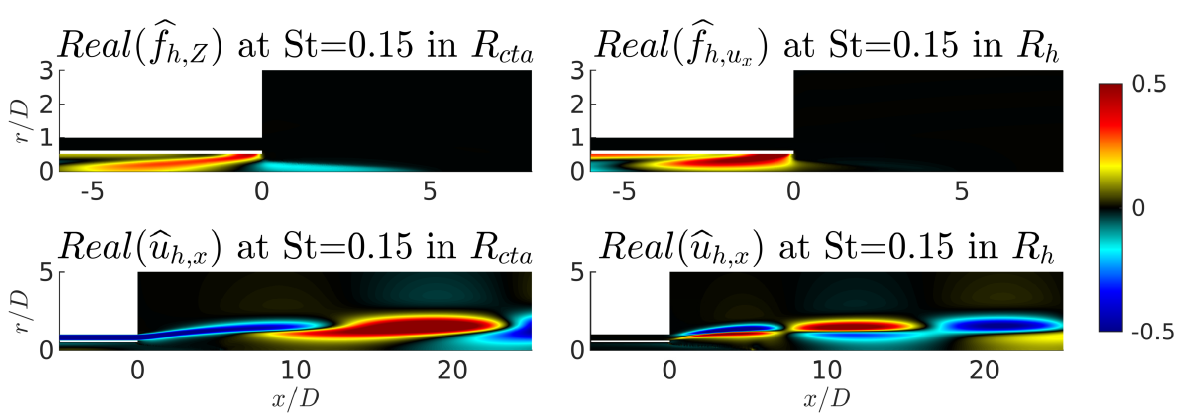}\label{compare_fz_p15}}%
\end{subfloatrow}}
\caption{Comparison between $\Rvec_{cta}$ and $\Rvec_{h}$. (a) Optimal forcing and response mode of $\Rvec_{cta}$ (left) and $\Rvec_{h}$ (right) with respect to $\widehat{\fvec}_{h,u_{x}}$ at $St=0.15$. (b) Optimal forcing and response mode of $\Rvec_{cta}$ (left) and $\Rvec_{h}$ (right) with respect to $\widehat{\fvec}_{h,Z}$ at $St=0.15$. All fields are normalized by their maximum value.}
\end{figure}

Besides the optimal gains, the corresponding forcing and response modes of $\Rvec_{cta}$ and $\Rvec_{h}$ with respect to $\widehat{\fvec}_{h}$ are compared. Here the frequency $St=0.15$ near the first strong peak value in the gain curves of $\Rvec_{cta}$ are taken as an example. The results are shown in the Fig.~\ref{compare_fu_p15} for $\widehat{\fvec}_{h,u_{x}}$ and Fig.~\ref{compare_fz_p15} for $\widehat{\fvec}_{h,Z}$. The response mode of hydrodynamic velocity $\widehat{\uvec}_h$ in $\Rvec_{cta}$ is caused by the linear superposition of the boundary condition due to the acoustic velocity $\widehat{\uvec}_a$ at the reference point and external forcing $\widehat{\fvec}_{h}$ in the flame domain $\Omega_{h}$. This result indicates that the acoustic-flame coupling terms in $\Rvec_{cta}$ could result in different flame responses that is ignored in $\Rvec_{h}$. One possible reason for the different behavior of $\Rvec_{cta}$ and $\Rvec_{h}$ is the weighting matrix $M_q$ used for output response measurement. The energy of acoustic pressure $\pvec_a$ is directly measured in $\Rvec_{cta}$, while only heat release rate fluctuation $\widehat{Q}$ in $\Rvec_{h}$. According to the Rayleigh criterion, the thermoacoustic system could become unstable when $\pvec_a$ and $\widehat{Q}$ are in the same phase, which could not be investigated with $\Rvec_{h}$. 

\begin{figure}
{\begin{subfloatrow}[1]
\sidesubfloat[]{\includegraphics[width=0.95\textwidth]{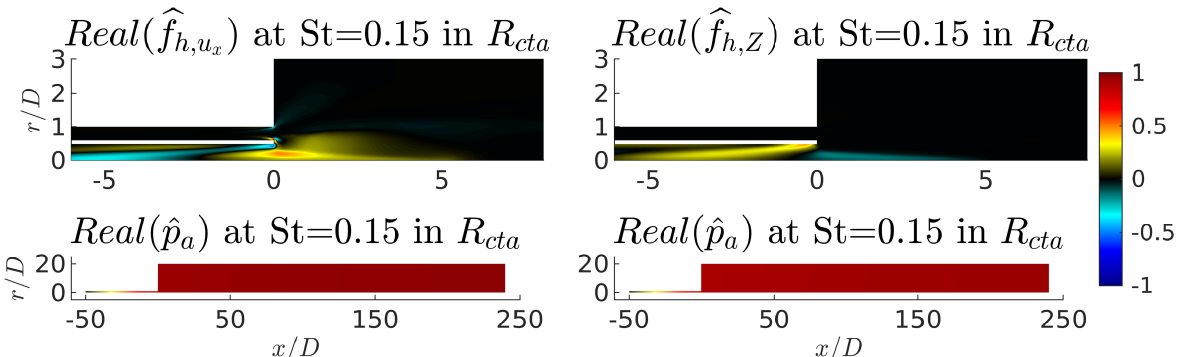}\label{fufz015}}%
\end{subfloatrow}}
{\begin{subfloatrow}[1]
\sidesubfloat[]{\includegraphics[width=0.95\textwidth]{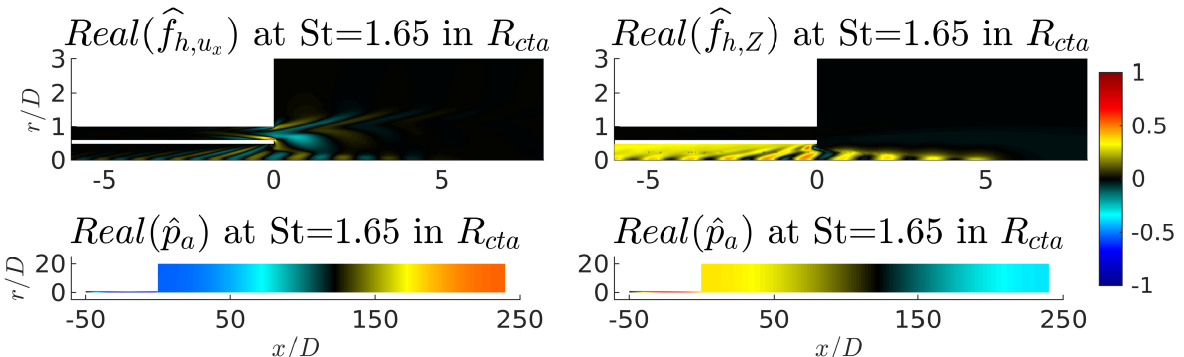}\label{fufz165}}%
\end{subfloatrow}}
\caption{Comparison between $\widehat{\fvec}_{h,u_{x}}$ and $\widehat{\fvec}_{h,Z}$ to $\Rvec_{cta}$. (a) Optimal forcing and response mode of $\widehat{\fvec}_{h,u_{x}}$ (left) and $\widehat{\fvec}_{h,Z}$ (right) to $\Rvec_{cta}$ at $St=0.15$. (b) Optimal forcing and response mode of $\widehat{\fvec}_{h,u_{x}}$ (left) and $\widehat{\fvec}_{h,Z}$ (right)  to $\Rvec_{cta}$ at $St=1.65$. All fields are normalized by their maximum value.}
\end{figure}

In addition to the different behaviour of $\Rvec_{cta}$ and $\Rvec_{h}$ with respect to hydrodynamic forcing $\widehat{\fvec}_{h}$, the roles of different forcing type $\widehat{\fvec}_{h}$ and $\widehat{\fvec}_{a}$ in $\Rvec_{cta}$ are investigated. Here $\widehat{\fvec}_{h}$ could be regarded as the forcing in the level of flame subdomain $\Omega_h$ and plays roles mainly near the flame front, while $\widehat{\fvec}_{h}$ in the level of whole combustor domain $\Omega$. From Fig.~\ref{norm}, the gain curve has many strong peak values. Here the frequencies $St=0.15$ and $St=1.65$ near the first two resonance eigenfrequencies are chosen for comparison. The optimal forcing and response modes of $\widehat{\fvec}_{h,u_{x}}$ and $\widehat{\fvec}_{h,Z}$ are shown in the Fig.~\ref{fufz015} and Fig.~\ref{fufz165}. 

\begin{figure}[htpb!]
\centering
\includegraphics[width=\textwidth]{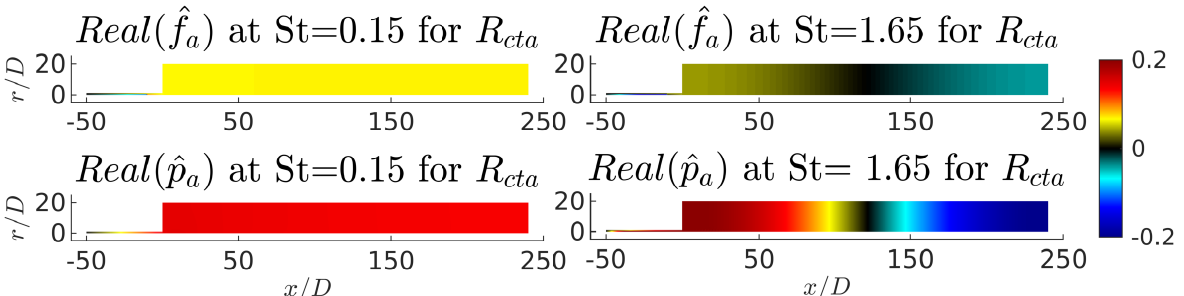}
\caption{Optimal forcing and response mode of $\Rvec_{cta}$ to $\widehat{\fvec}_{a}$ at $St=0.15$ (left) and $St=1.65$ (right). All fields are normalized by their maximum value.}
\label{c_fac_f}
\end{figure}

It is found that different hydrodynamic or compositional forcing in $\Omega_h$ can result in acoustic response with similar patterns but different amplitude amplification in $\Omega$. Compared to the baseline acoustic eigenmodes at the two frequencies, the response modes here could be regarded as response modes origin from perturbed acoustic eigenmodes. The corresponding results of $\widehat{\fvec}_{a}$ are shown in the Fig.~\ref{c_fac_f}. The forcing $\widehat{\fvec}_{a}$ is similar to the adjoint acoustic eigenmode in the previous research\cite{magri2013sensitivity}. The corresponding response modes resemble the acoustic fields induced by $\widehat{\fvec}_{h}$ but with different amplitude amplification.

\section{Conclusions}\addvspace{10pt}
A coupled thermoacoustic resolvent analysis framework is derived and applied to investigate the flame-acoustic interaction of a model two-stream coaxial combustor. Both the hydrodynamic and chamber acoustic effects are incorporated into the framework and coupled by matching the velocity disturbance at a reference location. The forcing-response relations of the flame-acoustic closed-loop system are constructed with the proposed resolvent method. It is found that the optimal gain corresponding to the coupled resolvent exhibits the peaks induced by pseudo-resonance amplification and acoustic resonance. In addition, the coupled resolvent allows us to evaluate the optimal forcing and response modes corresponding to different forcing mechanisms, including the hydrodynamic or compositional perturbations around the flame front and the acoustic forcing present in the whole chamber. It is recognized that the optimal response mode to the hydrodynamic perturbation preserves the rank-$1$ property, while the optimal response mode to acoustic forcing loses the rank-$1$ property at certain frequencies. Finally, it is proven that the adoption of optimized external input forcing can effectively cancel the acoustic pressure induced by other perturbations in the combustor. This showcases the potential of leveraging the derived resolvent framework to design active control strategies for thermoacoustic systems.

\acknowledgement{Declaration of competing interest} \addvspace{10pt}
The authors have no conflicts to disclose.

\acknowledgement{Acknowledgments} \addvspace{10pt}
This work is supported by the NSFC Basic Science Center Program for “Multiscale Problems in Nonlinear Mechanics” (No. 11988102).


 \footnotesize
 \baselineskip 9pt


\bibliographystyle{pci}
\bibliography{PCI_LaTeX}


\newpage

\small
\baselineskip 10pt



\end{document}